\documentclass[11pt]{article}

\usepackage{xcolor}
\usepackage{authblk}
\usepackage{graphicx}
\usepackage{cite}
\usepackage{amsmath}
\usepackage[left=3cm, right=3cm, top=3cm, bottom=4cm]{geometry}

\title{Direct Observation of a Feshbach-resonance by Coincidence-detection of Ions and Electrons in Penning Ionization Collisions}

\date{}
 
\newcommand\correspondingauthor{\thanks{email: edvardas.narevicius@weizmann.ac.il}}
\author[1]{Baruch Margulis}
\author[1]{Julia Narevicius}
\author[1]{Edvardas Narevicius\correspondingauthor}
\affil[1]{Department of Chemical and Biological Physics, Weizmann Institute of Science, Rehovot, Israel.}

\begin{document}

\maketitle

\begin{abstract} {
Observation of molecular dynamics with quantum state resolution is one of the major challenges in chemical physics. Complete characterization of collision dynamics leads to the microscopic understanding and unraveling of different quantum phenomena such as scattering resonances. Here we present an experimental approach for observing molecular dynamics involving neutral particles and ions that is capable in providing state-to-state mapping of the dynamics.
We use Penning ionization reaction between argon and metastable helium to generate argon ion and ground state helium atom pairs at  separation of several angstroms. The energy of ejected electron carries the information about the initial electronic state of  an ion. The coincidence detection of  ionic products provides a state resolved description of the post-ionization ion-neutral dynamics. We demonstrate that correlation between the electron and ion energy spectra enables us to directly observe the spin-orbit excited Feshbach resonance state of HeAr$^+$. We measure the lifetime of the quasi-bound HeAr$^+$  $A_\mathrm{2}$ state and discuss possible applications of our method.
 }  \end{abstract}

\section{Introduction}
The outcome of atomic and molecular collisions is governed by the details of the intermolecular interactions. In some cases, these can be accurately predicted and calculated by state-of-the-art quantum theory. In other instances, experiments serve as a direct probe of interactions. Starting with the pioneering work of Lee and Herschbach\cite{lee1969molecular}, gas phase collisions using adiabatically cooled molecular beams have become a central tool in chemical physics, allowing one to prepare reactants with a well-defined quantum state and collision energy\cite{lee1987molecular}.  
A complete state resolved detection of all reaction products however, is difficult in current experiments. Only recently Gao \textit{et al}.\cite{gao2018observation} reported a fully resolved detection of rotational excitation in bimolecular collisions.
The latest advances in molecular beams studies now enable performing precise experiments reaching collision energies as low as 10 mK, and the observation of different quantum phenomena such as scattering resonances\cite{henson2012observation}, the quantum isotope effect\cite{lavert2014observation}, stereodynamics of chemi-ionization reaction\cite{zou2019sub} and the prominent effect of the internal molecular structure on dynamics at low collision energies\cite{shagam2015molecular,klein2017directly}. 

Collisions are not the only option for observing molecular dynamics. A molecule can be excited into a non-stationary state in femtosecond transition-state spectroscopy where dynamical information can be recovered from a time-resolved spectroscopy of dissociation fragments \cite{zewail2000femtochemistry}. A non-stationary state can also be reached by a sudden quench of a system onto a lower electronic energy manifold. For example, slow photoelectron velocity-map imaging\cite{weichman2018slow} (SEVI) has been used to detect Feshbach resonances in the F+ H$_2$ reaction by measuring the photoelectron energy spectrum of the photodetachment of the FH$_2 ^-$ molecular anion  \cite{weichman2017feshbach}.

Here we present an experimental approach where ionization induced by collision between excited noble gas and a neutral particle is used to initiate dynamics between ions and noble gas atoms. We show that such a method is capable of quantum state-to-state resolution. Other experimental approaches for probing ion-neutral dynamics include crossed beam studies of a neutral and ionic beams\cite{mikosch2008imaging,stei2016influence}, collisions within the orbit of the Rydberg electron \cite{allmendinger2016new,allmendinger2016observation} and the co-trapping of ions and atoms\cite{hall2012millikelvin}. During Penning ionization (PI) reaction\cite{siska1993molecular}, sudden ionization of the neutral system takes place with the ejected electron serving as an indicator of the initial ionic quantum state. Following the interaction with the noble gas atom, Penning ions carry information about the final distribution of states. Therefore, the correlation between energy distributions of Penning electrons and Penning ions serves as a state-to-state mapping of the post-ionization ion-neutral dynamics.

A crucial part in our experiment is the coincidence detection of the momenta of Penning ionization charged products. This allows us to detect ion electron pairs that originate from the same collision event.  Coincidence detection of ionization fragments enabled many groundbreaking experiments including observation of the Efimov state of He trimer\cite{kunitski2015observation},  determination of chirality of molecules \cite{pitzer2017determine} and other experiments providing the insight the dynamic of photoionization \cite{dorner2000cold,ullrich2003recoil,eland1972photoelectron,sztaray2017crf}.
Many experiments reported separate measurement of electron and ion energy spectrum for several PI systems\cite{siska1993molecular,hotop1974analyses}. However, without the correlation, electron energy provides information only about the initial state of the PI ion. As such, it is similar to photoionization spectra and does not necessarily carry additional information. In contrast, the ion energy spectrum does provide information about nuclear dynamics however, it is always averaged over many initial states that are occupied during the PI process. The correlation is critical in resolving nuclear dynamics for a single initial state. Interestingly, this allows us to achieve quantum state-to-state resolution without any initial state preparation step that requires single quantum level control. 

As a first application of our method we demonstrate a direct detection of Feshbach HeAr$^+$  molecule that dissociates on a microsecond timescale due to the spin-orbit coupling.
In future, we plan to apply this method to investigate post-PI nuclear dynamics between vibrationally excited H$_2 ^+$ molecular ion and helium atom. We will be able to observe state-to-state inelastic processes as well as reactive collisions leading to the formation of HeH$^+$ molecule.

\section{Results}

\subsection{Dynamics of Penning ionization}

\begin{figure}[t]
\centering
\includegraphics[width=0.9\textwidth]{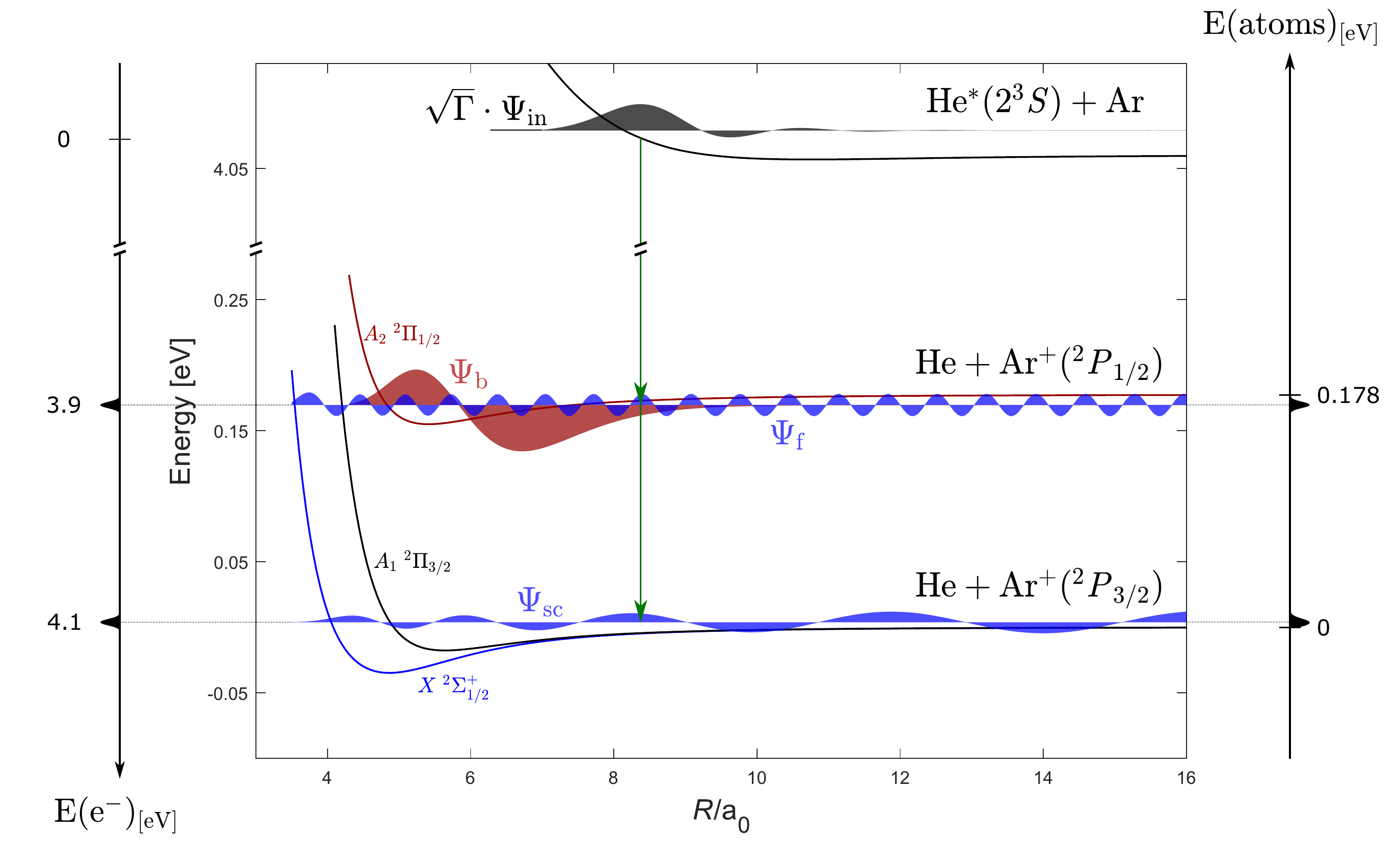}
\caption{ Potential energy curves, obtained from\cite{hapka2013first,carrington1995microwave}. $\Psi \mathrm{_{in}}$ represents the 
incoming wave function with wave number matching an energy of $k\mathrm{_B}\times$220K. $\Psi \mathrm{_{b}}$ represents the first excited vibrational level of $\mathrm{HeAr^+}$ at the $A\mathrm{_2}$ electronic state. $\Psi \mathrm{_{f}}$ represents the scattering state wave function of the $X$ state which couples to $\Psi \mathrm{ _{b}}$.  $\Psi \mathrm{_{sc}}$ represents the scattering state wave function of the $X$ state with maximal overlap with $\sqrt{\Gamma}\cdot \Psi \mathrm{_{in}}$.  E(e$^-$) represents energy distribution of ejected electrons and E(atoms) represents the energy distributed between He and Ar$^+\mathrm{(^2 {\it P}_{3/2})}$.
 All wave functions were calculated using discrete variable representation \cite{colbert1992novel}, the resonance width $\Gamma(R)$ is taken from \cite{bibelnik2019cold}.}
 \label{potentials}
\end{figure}

As a first demonstration of our approach we study the PI and associative ionization (AI) channels of the collision between metastable helium and neutral argon:

\begin{equation}
{\rm He^*(2^1 {\it S},2^3 {\it S})+Ar}\rightarrow \begin{cases}
{\rm He+Ar^+(^2 {\it P}_{1/2},^2 {\it P}_{3/2}) + e^-} &  \text{PI} \\
{\rm HeAr^+({\it X,A}_1,{\it A}_2) + e^- }& \text{AI}
\end{cases}
\end{equation}

A schematic of the relevant potential curves together with calculated wave functions is presented in Fig. \ref{potentials}.
The ionization process is described by the projection of the incoming wave function weighted by the square root of the autoionization width $\Gamma(R)$ on one of the three possible ion-neutral potential surfaces, which are asymptotically separated on energy scale by the spin-orbit interaction energy of Ar$^+$ ($\mathrm{\Delta E _{SO}}$ = 0.178 eV). The energy of the ejected electron corresponds to the difference between the neutral and ionic potential curves \cite{miller1970theory}; therefore, information about the formation of a particular state is manifested in the electron energy distribution $\mathrm{E(e^-)}$. 
The projection on the manifold of ionic states may result in a Ar$^+$ product (PI), or a HeAr$^+$ product (AI).
However, a molecular ion generated in the electronically excited $A_\mathrm{2}$ state ($\Psi \mathrm{ _{b}}$) may decay via a Feshbach resonance\cite{feshbach1958unified,fano1961effects} to the continuum of the electronic ground $X$ state ($\Psi \mathrm{ _{f}}$). As a result of the predissociation, the free Ar ion and neutral He acquire kinetic energy on the order of $\mathrm{\Delta E _{SO}}$. 
 The kinetic energy distribution of Ar$^+$ ($\mathrm{E(atoms)}$) consists of low kinetic energy ions resulting from the direct PI reaction and high kinetic energy ions resulting from the predissociation process. In the following section we present how the correlation between $\mathrm{E(e^-)}$ and $\mathrm{E(atoms)}$ provides the observation of the process, by relating the high kinetic energy argon ions to a specific electron energy.
 
\subsection{Observation of the scattering resonance}

\begin{figure}[t]
\centering
\includegraphics[width=0.7\textwidth]{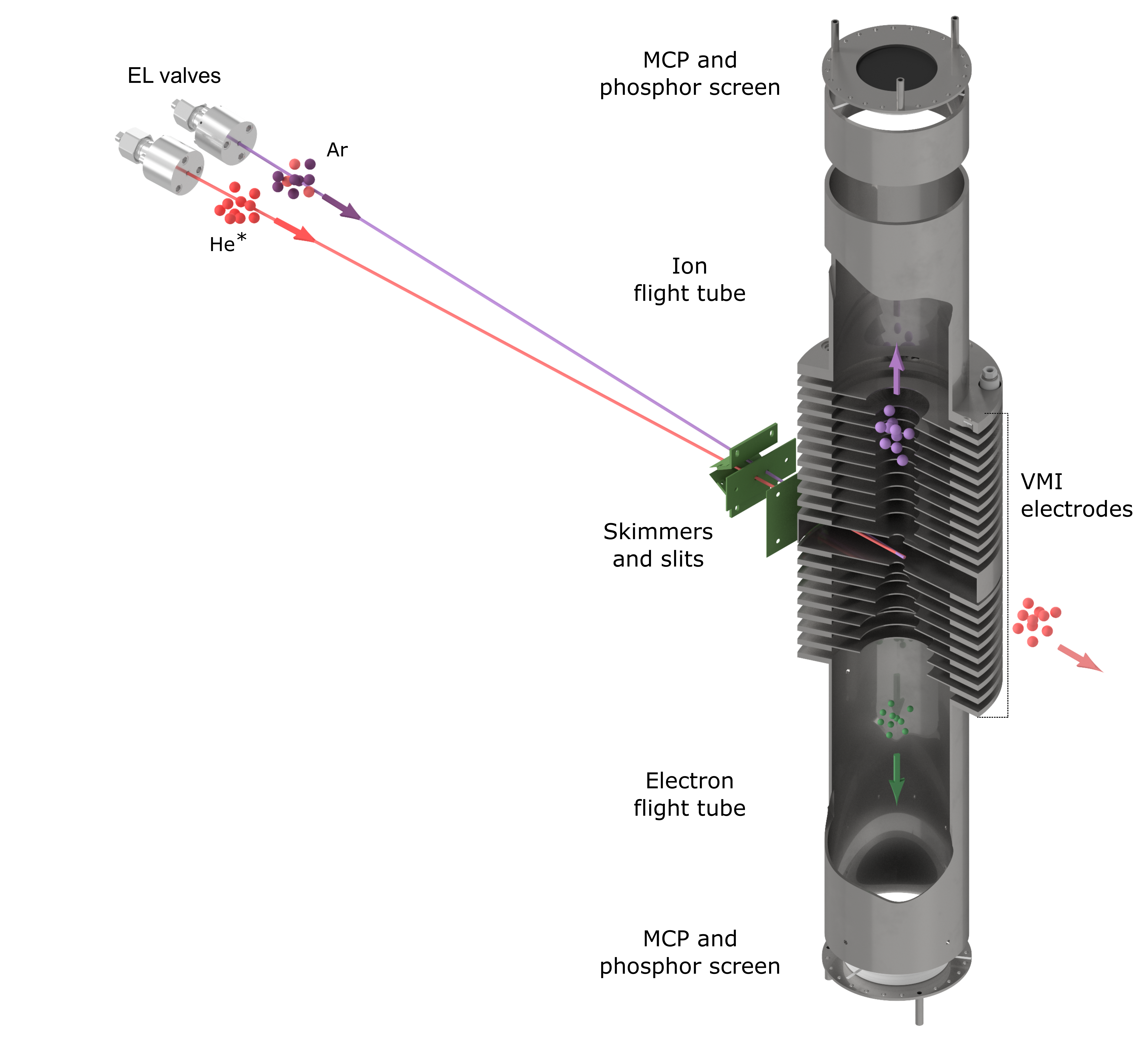}
\caption{Experimental setup. Supersonic beams of metastable He (red) and ground state argon (Purple) are shaped by a set of skimmers and slits before colliding in the center of a double VMI apparatus. Product ions and electrons are accelerated and their velocity is imaged by a set of 19 open-aperture electrodes followed by two free-field flight tubes. For both ions and electrons, arrival times and positions are detected by micro channel plate (MCP) detectors followed by phosphor screens.}
\label{exp_set}
\end{figure}

For the purpose of this study we designed and constructed a Coincidence Double Velocity Map Imaging (CDVMI) apparatus. A schematic of the setup is presented in Fig. \ref{exp_set}.
The 2D velocity distribution and the time-of-flight (TOF) of ions and electrons are obtained by a two-sided VMI \cite{eppink1997velocity} spectrometer.
Coincidence ion-electron pairs for a specific ion mass were selected by their TOF (cf. Methods) and their arrival positions were identified by correlation between optical brightness and electronic signal magnitude\cite{urbain2015zero}.
Mass-specific VMI images of ions and corresponding electrons are produced with complete mapping between the images. 

\begin{figure}[t]
\centering
\includegraphics[width=0.9\textwidth]{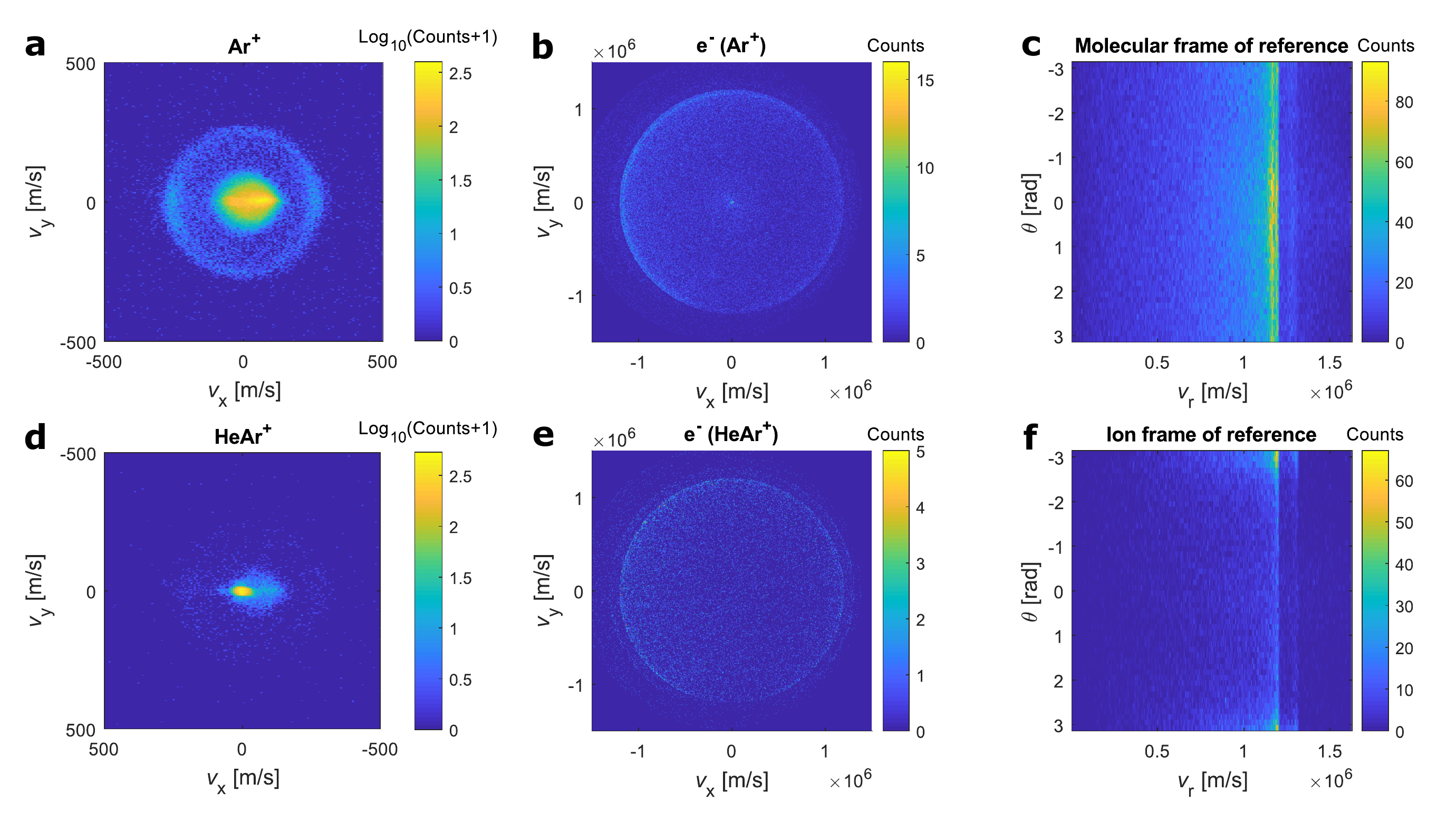}
\caption{Coincidence VMI images of Ar$^+$ and HeAr$^+$ (a,d) and coincidence electrons (b,e). Ionic data is presented on a log scale to emphasize the outer circular feature which is the main interest of this paper. 
c. and f. are electron VMI image after rotation to the ionic frame of reference, where $\theta$ is defined as the relative angle between the 2D velocities of the electron and the ion.}
\label{CVMI_ar_hear}
\end{figure}

Mass-selected VMI images of Ar$^+$, HeAr$^+$ and coincidence electrons are presented in Fig. \ref{CVMI_ar_hear}. The VMI image of Ar$^+$ consists of a central feature which presents low kinetic energy, forward-scattered Ar$^+$ produced directly by PI, and a circular feature related to high kinetic energy Ar$^+$, with a velocity magnitude centered at 269 m/s with a width of 16 m/s (all widths reported in this manuscript refer to half width at half maximum of a Gaussian distribution).  
The velocity magnitude of high kinetic energy Ar$^+$ corresponds to a kinetic energy of 0.164 $\pm$ 0.019 eV which is partitioned between Ar$^+$ and neutral He. 
For HeAr$^+$, the ionic VMI image consists of a single central feature at a center-of-mass velocity with an width of 19 m/s and 13 m/s in longitudinal and transverse directions.

\begin{figure}[t]
\centering
\includegraphics[width=0.75\textwidth]{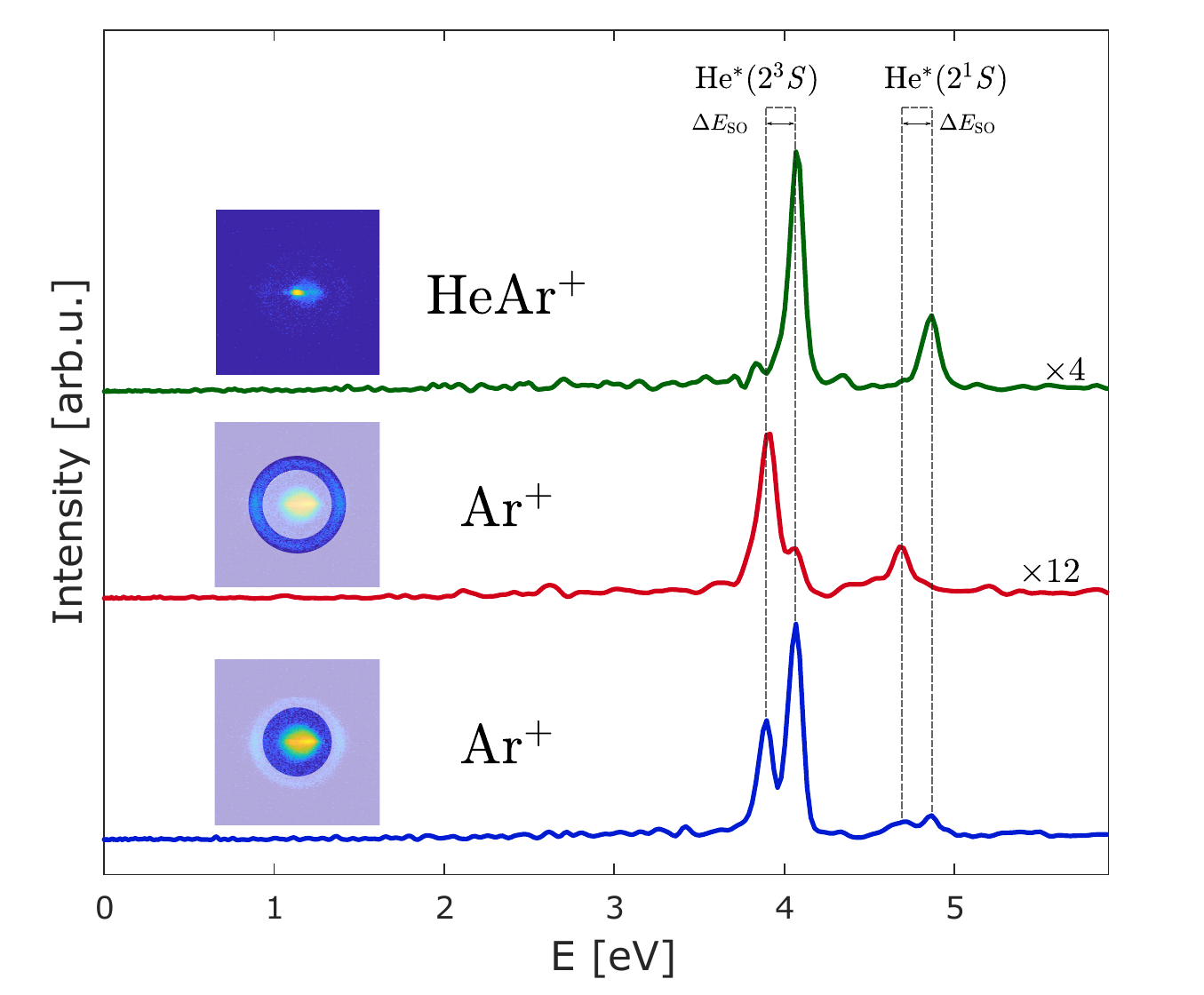}
\caption{Energy distribution of Penning electrons related to Ar$^+$ with radial velocity between 0 and 209 m/s (blue curve), Ar$^+$ with radial velocity between 209 and 295 m/s (red curve), and HeAr$^+$ (green curve). The related ionic VMI images are presented above each plot.}
\label{el spec}
\end{figure}

The complete mapping between the ionic and the electronic VMI images enables the reconstruction of electronic VMI images corresponding to a given range of ionic radial velocities. Energy distribution of electrons were obtained by inversion of the rotated electron VMI images using MEVELER \cite{dick2014inverting} (cf. Methods).
The electron energy distributions related to low and high kinetic energy Ar ions is presented in Fig. \ref{el spec} as blue and red curves.  
The electron energy distribution related to low kinetic energy Ar ions contains two main peaks at 3.89 and 4.06 eV arising from the ionization of Ar by He$^*$($\mathrm{^3} S$). The two smaller peaks at around 4.87 and 4.69 eV  correspond to ionization of Ar by He$^*$($\mathrm{^1} S$).
The electron energy distribution related to high kinetic energy Ar ions presents a clear preference for electron energies which pinpoints the origin of those ions as the $A_\mathrm{2}$ electronic state. The observed additional kinetic energy of $\sim$0.164 eV is gained by resonantly decaying from the quasi-bound $A_\mathrm{2}$ state to the free particle Ar${\rm ^+ ( ^2 {\it P}_{1/2} )}$+He state as illustrated in Fig. \ref{potentials}.

The complementary dynamical picture is given by inspection of the AI channel. The electron energy distribution for electrons related to HeAr$^+$ is presented in Fig. \ref{el spec} (green curve) and shows clear preference for electron energies representing the formation of molecular ions at the $X$ and $A_\mathrm{1}$ states. Any quasi-bound state formed upon ionization at the electronic $A_\mathrm{2}$ state dissociates with a lifetime much shorter than the particle TOF, and therefore is undetected as a molecular ion.

\begin{figure}[t]
\centering
\includegraphics[width=0.5\textwidth]{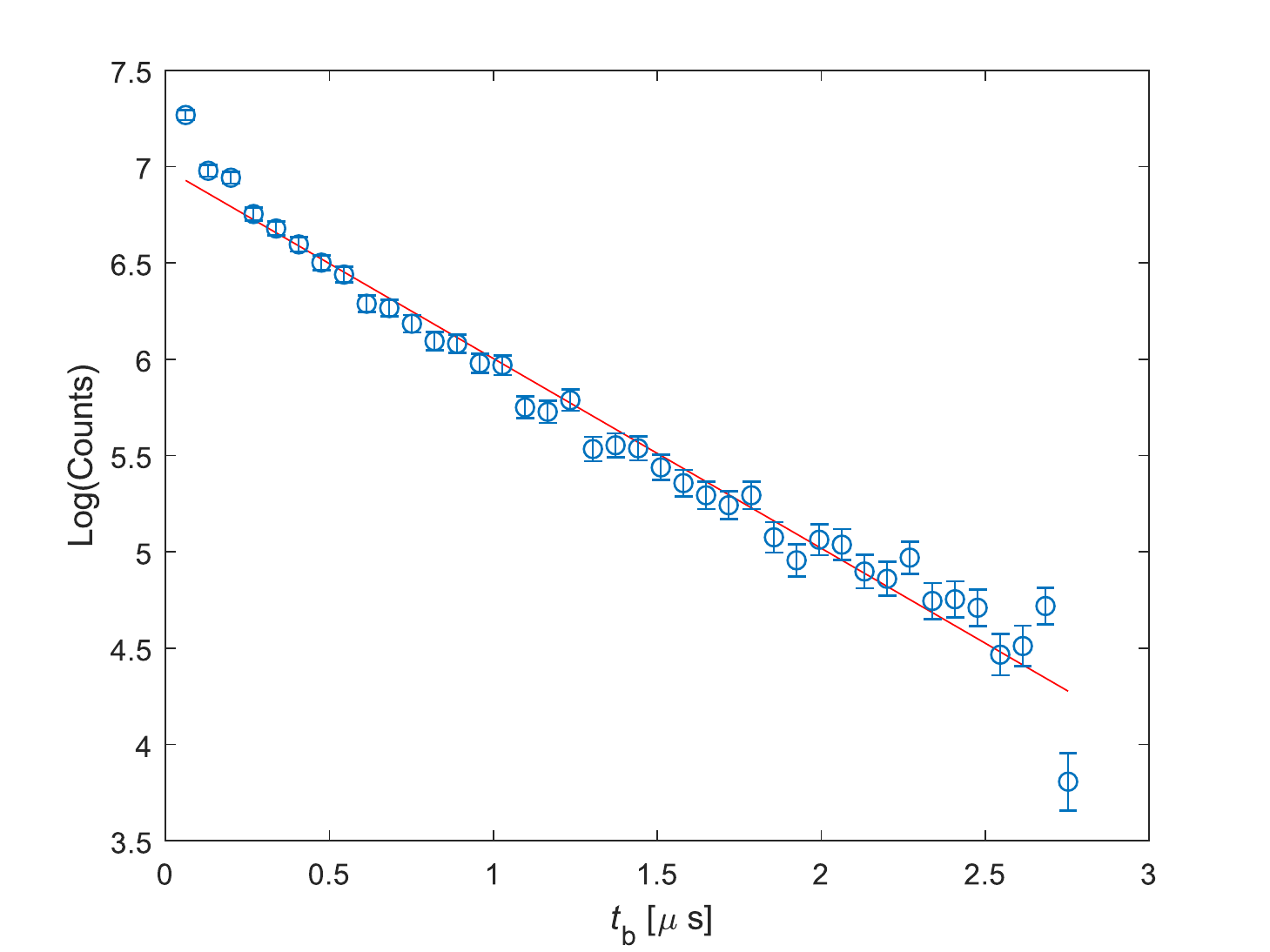}
\caption{Lifetime of the quasi-bound $\mathrm{HeAr^+}$ $A_\mathrm{2}$ state calculated from the positive addition to the TOF of Feshbach Ar ions. Error bars represent statistical errors. The red line represents an exponential fit. }
\label{Life time}
\end{figure}

Free argon ions which originate from the dissociation of the quasi-bound ArHe$^+$ $A_\mathrm{2}$ state spend a fraction of their trajectory as a particle with increased mass, which results in a finite addition to their TOF. The lifetime of the quasi-bound $A_\mathrm{2}$ state can be extracted from the TOF of related ions using 
\begin{equation} 
t_\mathrm{b} = \Delta\mathrm{TOF} /(1-\sqrt{\frac{m_{\mathrm{Ar}}}{m_\mathrm{HeAr}}})
\end{equation}
 where $t_\mathrm{b}$  is the absolute TOF of the quasi-bound HeAr$^+$  in the  $A_\mathrm{2}$ state before dissociation,  $\Delta\mathrm{TOF}$ is the addition to the ions TOF relative to the center TOF peak of Ar$^+$ and $\frac{m_{\mathrm{Ar}}}{m_\mathrm{HeAr}}$ is the mass ratio.
Fig. \ref{Life time} presents the observed exponential decay of $t_\mathrm{b}$ for Ar$^+$ with a radial velocity between 209 and 295 m/s, with an observed lifetime of 1.01 $\mathrm{\mu s}$.
Our lifetime measurement is not sensitive enough to resolve lifetimes arising from different rovibrational states, however,  the measured lifetime falls into the range predicted by theory\cite{blech2020phase}. Further support of our observation is given by the lack of Feshbach ions in previous studies of Ne$^*$+Ar PI reaction due to the long lifetime of the quasi-bound NeAr$^+$ $A_\mathrm{2}$ state\cite{blech2020phase,delmdahl2000crossed}.

In conclusion, we have presented an experimental method which provided a state resolved description of ion-neutral dynamics. For the He-Ar$^+$ system this enabled the observation of a Feshbach resonance identified as the predissociation of the HeAr$^+$ $A_\mathrm{2}$ molecular state.
Although the kinetic energy distributions of He$^*$+Ar PI products were obtained in previous experiments  \cite{blech2020phase,ohno1996collision,longley1993recent}, the lack of correlation between ionic and electronic data left the details of the post-ionization dynamics completely hidden. Coincidence detection of ions and electrons provides the mapping between initial and final states and enables the direct observation of the scattering resonance.

Our approach may be generalized to any ion-neutral system accessible via PI reaction.
Specifically, study of PI of molecules, where rovibrational degrees of freedom span the manifold of ion-neutral potential curves, may pave the way to a multitude of possible experiments studying state-to-state inelastic and reactive collisions between molecular ions and neutral atoms. Specifically, the study of PI ionization of hydrogen molecule by metstable helium will enable the observation of state-to-state collisional quenching of vibrationally excited H$_2 ^+$ and state-to-state reactive H$_2 ^+$+He collisions leading to the formation of HeH$^+$.

\newpage
\section{Methods}

\subsection{Experimental setup:}

A pulsed supersonic beam of a 50/50\% mixture of Argon and Helium was created by a Even-Lavie (EL) valve \cite{even2000cooling} and had a mean velocity of 784 m/s and a width of  21 m/s. A second pulsed beam of pure Helium was created by another EL valve oriented at an angle of $4.5^0$ relative to the first valve and had a mean velocity of 1786 m/s and a width of 72 m/s. Helium was excited to the ${\rm 2^1}S$ and ${\rm 2^3} S$ states by a dielectric barrier discharge \cite{luria2009dielectric} located directly after the valve.
The beams passed through a 1 mm skimmer located 43 cm downstream from the valves orifice. The beams were then further shaped by a set of two double circular slits with diameter of 3 mm and 1 mm located 2.5 cm and 6.5 cm after the skimmer accordingly.  
The two beams crossed at the center of the CDVMI apparatus 58 cm relative to the valves orifice, where atoms collided with a collision energy of $\mathrm{k_B}\times$ 222 $\pm$ 33 K.
Ions and electrons were accelerated and velocity-imaged by a set of 19 electrodes, 8 on the electron side and 11 on the ion side. The electrodes had varying inner diameter (10-48 mm) and fixed outer diameter of 130 mm and were spaced by a 10 mm distance.
Following the VMI electrodes, particles entered a field free region defined by two flight tubes, 20 cm long on the electron side and 62 cm long on the ion side. 
The electron flight region is surrounded by a cylindrical sheet of mu-metal for protection against the effect of external magnetic fields. 
Both ions and electrons were accelerated to an energy of about 2 keV towards a time and position sensitive MCP detector followed by a phosphor screen, P46 on ion side and P43 on electron side. The potentials on all electrodes and MCP plates were kept constant throughout the experiment.
Arrival positions of ions and electrons were captured by a CCD camera and their arrival times were digitized using a fast scope. 
Experiments were preformed in a 10Hz repetition rate. Results presented in this work were obtained from a data set containing 1.1 million valves shots.
The exact geometry of the ion optics was designed based on classical trajectory simulations using SIMION program, optimized for maximal resolution $\frac{v}{\Delta v}$. 
For HeAr ions at 875 m/s SIMION simulations predict VMI resolution of 3.2 m/s. Experimental electron-recoil corrected VMI image of HeAr ions demonstrated transverse width of 4.1 m/s which corresponds to energy resolution of 0.004 meV. The resolution along the beam propagation axis is limited by the spread in center-of-mass velocity ($\sim$20 m/s). For 4.06 eV electrons, the observed width of 49 meV provides the upper limit for the resolution which is predicted by SIMION simulations to be 25 meV.

\subsection{ Data analysis and coincidence }

For every valve shot, the acquired experimental data included ions and electrons arrival times and amplitude of electronic signal as measured on the MCP back plate, arrival position and the magnitude of the optical signal as obtained from the recorded images of the phosphorus detector.
The mass distribution of detected ions was obtained by plotting a histogram of their TOF (cf Mass Spectrum). 
The obtained mass spectrum provides the limits of ion-electron TOF difference which are used for identification of coincidence in certain mass-to-charge ratio ion electron pairs. Coincidence by TOF is possible in our experiment due to the narrow spread in the TOF for a given ion mas- to-charge ratio ($\sim$0.01 $\mathrm{\mu s}$) relative to the large overlap in time between the reactant beams ($\sim$30 $\mathrm{\mu s}$). Once an ion-electron pair is identified by TOF, the related ion and electron arrival positions are selected from the measured optical signals based on the correlation between optical phosphorus detector brightness and amplitude of electronic MCP signal \cite{urbain2015zero}.
The resulted mass selected coincidence VMI images of ions and electrons are shown in Fig \ref{CVMI_ar_hear}. 
The ratio between camera pixels and velocity on the ionic VMI images was found by examining the shift in position of center of mass velocity for a variety of single beam Ar-He gas mixtures.

\subsection{Mass spectrum:}
\begin{figure}
\centering
\includegraphics[width=0.7\textwidth]{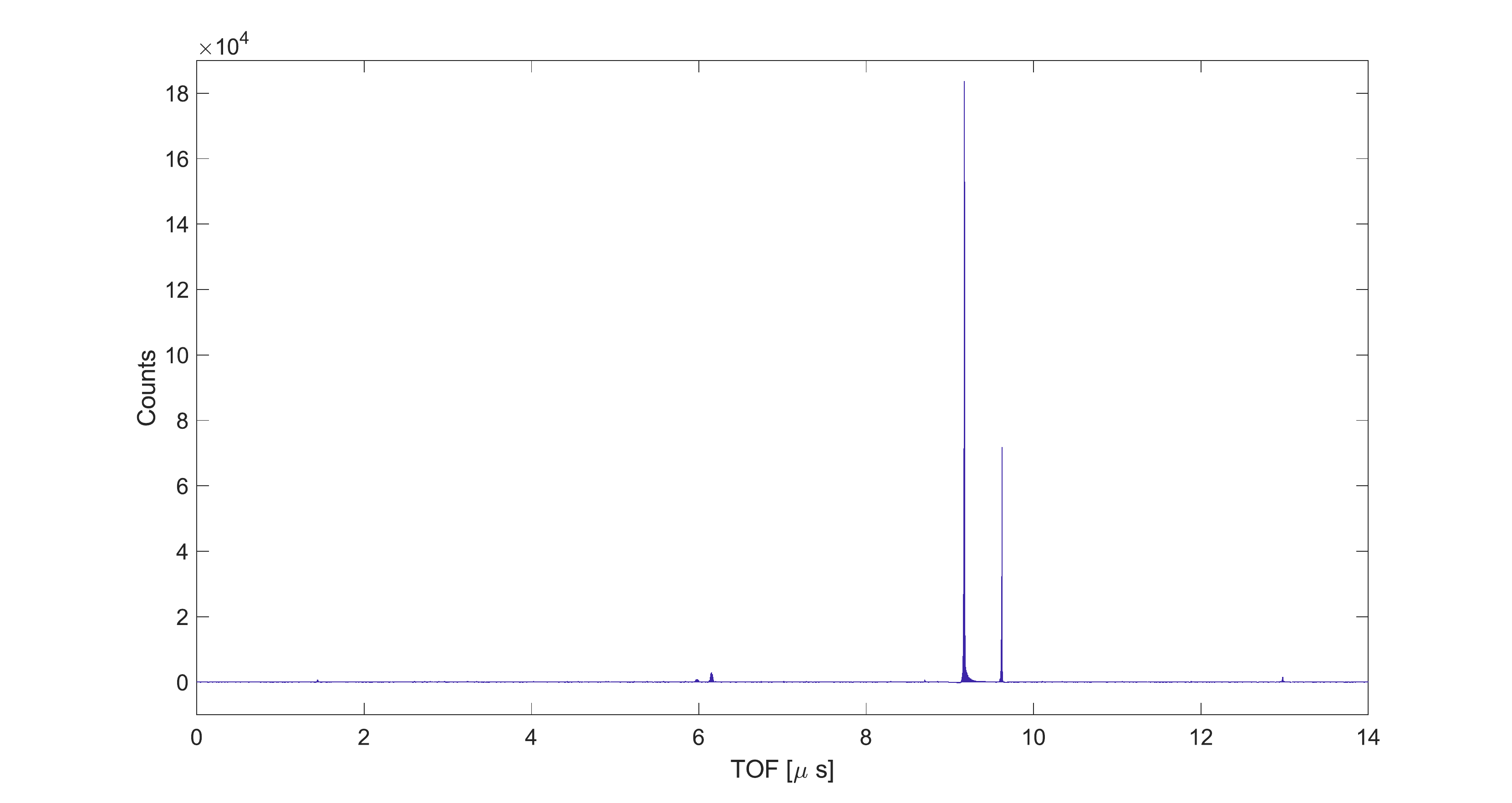}
\caption{TOF histogram of detected ions. Zero time stamp is defined by arrival time of electrons. The two main peaks correspond to  Ar$^+$ and HeAr$^+$. The two smaller peaks at $6.0 \mu s$ and $6.2 \mu s$ correspond to TOF of $\mathrm{H_2 O}$ and OH accordingly. The peak at $13 \mu s$ corresponds to TOF of $\mathrm{Ar_2}$.}
\label{MS}
\end{figure}

TOF distribution of detected ions is presented at Fig. \ref{MS}. We observe two main peaks at TOF of $9.2 \mu s$ and $9.6 \mu s $ which correspond to Ar$^+$ and HeAr$^+$ accordingly. CDVMI performs as a mass spectrometer with a resolving power (FWHM) of $\frac{m}{\Delta m}=1000$ for Ar and  $\frac{m}{\Delta m}= 1500$ for HeAr.
 The use of CDVMI as a mass spectrometer may be applied to any sample containing atoms and molecules with ionization energy smaller then excitation energy of the metastable atom.

\subsection{Electron energy distributions}

Information about energy distribution of electrons requires the inversion of the 2-D velocity distribution (Fig. \ref{CVMI_ar_hear} b. and e.) to a 3-D velocity distribution \cite{ashfold2006imaging}.
Since any inversion method requires an axis of cylindrical symmetry, the electron images were rotated to the ionic frame of reference (Fig. \ref{CVMI_ar_hear} c. and f.). We attribute the anisotropy of the electron VMI images to a non uniform collection efficiency, resulting from the mismatch of beams overlap position relative to the VMI orifice. 
Electron energy distributions presented in Fig. \ref{el spec} were obtained by inversion of rotated electron VMI images using MEVELER \cite{dick2014inverting}.
The resulted velocity magnitude distribution was scaled to energy scale by the known energy difference between metastable states of He.

\section{Acknowledgments} We thank H. Sade of the Weizmann CNC Section and A. Kuprienko of the Weizmann Chemical Research Support for assistance in designing and manufacturing the experiment components. We acknowledge funding from the European Research Council and the Israel Science Foundation.

\section{Data availability}
All relevant data are available from the authors upon reasonable request.

\section{Author Contributions}
B.M., J.N. and E.N. designed and constructed the CDVMI apparatus. B.M. performed the measurements. E.N. planned and supervised the project.

\section{Competing Interests}
The authors declare no competing interests.

\newpage
\bibliographystyle{ieeetr}
\bibliography{refs}

\end{document}